\documentclass[submission,copyright,creativecommons]{eptcs}
\providecommand{\event}{QPL 2022}
\pdfoutput=1

\newcommand{\Z}{\text{Z}}
\newcommand{\X}{\text{X}}

\newcommand{\I}{\text{I}}
\newcommand{\cxcount}[1]{\text{count}_\text{CX}\left(#1\right)}
\newcommand{\cxconj}[2]{#1\left(#2\right)}

\usepackage[utf8x]{inputenc}
\usepackage[greek, english]{babel}
\usepackage{textgreek}
\usepackage{amsmath}
\usepackage{hyperref}
\usepackage[numbers,sort&compress]{natbib}
\usepackage[inkscapelatex=false]{svg}

\usepackage[frozencache]{minted}     

\title{Annealing Optimisation of Mixed ZX Phase Circuits}
\author{
    Stefano Gogioso
    \institute{Hashberg Ltd}
    \email{quantum@hashberg.io}
    \and
    Richie Yeung
    \institute{Quantinuum Ltd}
    \email{richie.yeung@cambridgequantum.com}
}
\def\titlerunning{Annealing Optimisation of Mixed ZX Phase Circuits}
\def\authorrunning{S. Gogioso and R. Yeung}

\begin{document}

\maketitle

\begin{abstract}
We present a topology-aware optimisation technique for circuits of mixed ZX phase gadgets, based on conjugation by CX gates and simulated annealing.
\end{abstract}

\section{Introduction}

In this work, we build upon the Master's thesis by one of the authors \cite{yeung2020diagrammatic} and present a topology-aware optimisation technique for circuits of mixed ZX phase gadgets, based on conjugation by CX gates and simulated annealing.
The basic rules on CX conjugation of phase gadgets have previously appeared in the literature \cite{kissinger2020reducing,cowtan2019phase,cowtan2020generic,de2020architecture,nash2020quantum}---which features other topology-aware techniques---and are used by both the PyZX library \cite{kissinger2019pyzx,pyzx2019} and the t$|$ket$\rangle$ compiler \cite{sivarajah2020t,pytket2019}.
We test the performance of our optimisation technique on random circuits of mixed ZX phase gadgets.
An open-source implementation is made available as part of the Python library \texttt{pauliopt} \cite{pauliopt2021}, which is also used to generate the circuit figures in this work.

With NISQ applications in mind, our optimisation target is the number $\cxcount{P}$ of nearest-neighbour (NN) CX gates required to implement a mixed ZX phase circuit $P$ on a given topology, making the following assumptions on compilation:
\begin{itemize}
  \item Each phase gadget will be compiled to a single-qubit rotation conjugated by trees of CX gates.
  \item Long-range CX gates in the topology will be compiled to double-ladders of NN CX gates.
\end{itemize}
In these circumstances, the CX count of an individual phase gadget depends on the mutual distances between its legs.
Specifically, it is computed by finding a minimum spanning tree with weights related to the distance of leg qubits in the given topology.

Our optimisation technique is based on the observation that conjugating a given mixed ZX phase circuit $P$ with an arbitrary CX circuit $C$ has the effect of changing the legs of the individual gadgets, without otherwise altering their angles or position within the circuit.
The resulting circuit $\cxconj{C^\dagger}{P} := C^\dagger \circ P \circ C$ is again a mixed ZX phase circuit, related to the original circuit by:
\[
P = C \circ \cxconj{C^\dagger}{P} \circ C^\dagger
\]
However, the different gadget legs in $\cxconj{C^\dagger}{P}$ might result in a lower overall CX count:
\[
2\cdot\cxcount{C}+\cxcount{\cxconj{C^\dagger}{P}}
\]
Finding a CX circuit $C_{opt}$ which (approximately) minimises the overall CX count above results in an optimised implementation of our initial mixed ZX phase circuit $P$, expressed as conjugation of another mixed ZX phase circuit $\cxconj{C_{opt}^\dagger}{P}$ by the CX circuit $C_{opt}$:
\[
P=C_{opt} \circ \cxconj{C_{opt}^\dagger}{P} \circ C_{opt}^\dagger
\]
We use simulated annealing to explore the space of conjugating CX circuits and find an approximately optimal $C_{opt}$, starting from the empty CX circuit.
We fix a number of layers and explore the space by progressively introducing or removing NN CX gates within those layers, altering the overall CX count in the process.
In the early stages of optimisation, the annealing ``temperature'' $t$ is high and we are free to explore the configuration space, even at the expense of a---hopefully temporary---increase in overall CX count.
As the optimisation progresses, the temperature $t$ is lowered and the search becomes progressively more greedy, accepting changes which increase the CX count by $\Delta$ with progressively lower probability:
\[
\text{Prob}\left(\text{accept CX count increase $\Delta$}\right)
=
\dfrac{1}{2^{\Delta/t}}
\]
Our technique is applicable to parametric circuits, such as the ansatzes used in quantum machine learning, adiabatic quantum computation \cite{farhi2000quantum} and quantum approximate optimisation \cite{farhi2014quantum}.
Such ansatzes typically consist of a large number of repeating layers, using the same phase gadgets with possibly different parameters.
In this context, a key observation about our technique is that a single conjugating CX circuit is sufficient to optimise all layers at once.
The number $K$ of layers can then be absorbed into the overall CX count calculation, with a single layer $L$ being optimised:
\[
2\cdot\cxcount{C}+K\cdot\cxcount{\cxconj{C^\dagger}{L}}
\]
As a consequence, the optimisation cost scales with the size of $L$, regardless of $K$, making our technique especially suited for application in the aforementioned domains.

\section{ZX Phase Gadgets}

A \emph{Z phase gadget} on $n$ qubits is the exponential of an imaginary scalar multiple of an element of the Pauli group $P_n$ on $n$ qubits involving only the $\Z$ and $\I$ Pauli matrices (and with scalar factor $+1$):
\[
  \exp\left(i \theta \bigotimes_{q=0}^{n-1} G_q\right)
  \hspace{5mm}\text{where}\hspace{5mm}
  \begin{cases}
    G_q = \Z &\text{ if qubit $q$ is a leg}\\
    G_q = \I &\text{ otherwise}
  \end{cases}
\]
The element $\bigotimes_{q=0}^{n-1} G_q \in P_n$ is known as the \emph{generator}, the parameter $\theta \in U(1)$ is known as the \emph{angle}, and the qubits $q$ where $G_q = \Z$ are known as the \emph{legs} of the gadget.
An \emph{X phase gadget} on $n$ qubits is analogously defined, using the Pauli matrix $\X$ in place of $\Z$.
Below are two examples of phase gadgets on 3 qubits:
\[
  \exp\left(i\frac{\pi}{4} \Z \otimes \I \otimes \Z \right)
  \hspace{3cm}
  \exp\left(i\frac{3\pi}{4} \X \otimes \X \otimes \X \right)
\]
Phase gadgets can be represented in the ZX calculus by attaching spiders for the gadget basis to each leg and joining them at a ``hub'', formed by spiders of alternating colours and including the angle ($\theta = \frac{\pi}{4}$ and $\theta = \frac{3\pi}{4}$ respectively) as a spider phase.
Below are the same two examples represented in the ZX calculus (Z phase gadget on the left, X phase gadget on the right):
\[
  \includegraphics[height=0.1\textwidth]{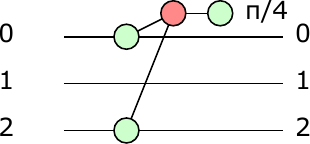}
  \hspace{3cm}
  \includegraphics[height=0.1\textwidth]{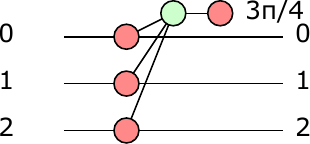}
\]

\subsection{Conjugation by CX gates}

Z and X phase gadgets are particularly well-behaved under conjugation by CX gates: they either keep the same legs, they gain a leg, or they lose a leg, but basis and angle are always left unchanged.
Instead of thinking of CX gates in terms of \emph{control} and \emph{target}, we think of them as having a Z qubit (the control) and an X qubit (the target).
Z phase gadgets are left unchanged under conjugation if the X qubit is not a leg:
\[
  \includegraphics[height=0.08\textwidth]{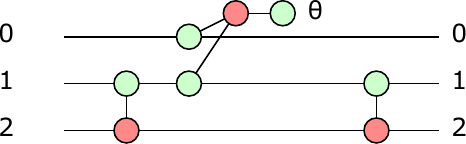}
  \hspace{1cm} \raisebox{0.03\textwidth}{=} \hspace{1cm}
  \includegraphics[height=0.08\textwidth]{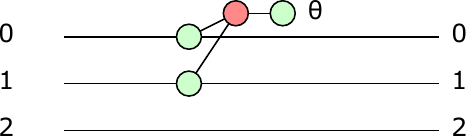}
\]
If the X qubit is a leg and the Z qubit is not a leg before conjugation, the Z qubit becomes a leg after conjugation:
\[
  \includegraphics[height=0.08\textwidth]{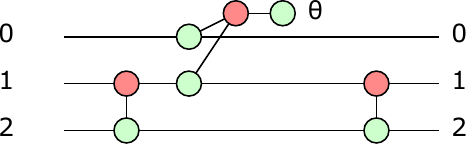}
  \hspace{1cm} \raisebox{0.03\textwidth}{=} \hspace{1cm}
  \includegraphics[height=0.08\textwidth]{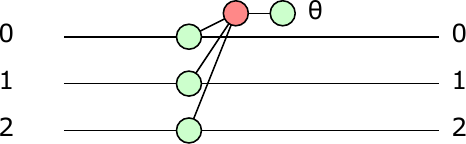}
\]
If the X qubit is a leg and the Z qubit is a leg before conjugation, the Z qubit is no longer a leg after conjugation:
\[
  \includegraphics[height=0.08\textwidth]{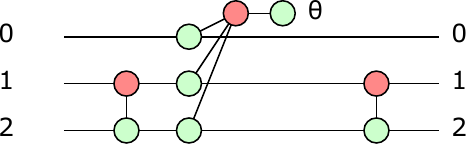}
  \hspace{1cm} \raisebox{0.03\textwidth}{=} \hspace{1cm}
  \includegraphics[height=0.08\textwidth]{svg-inkscape/gadg-cx-flip-example-1_svg-raw.pdf}
\]
The behaviour of X phase gadgets is obtained by inverting Z and X everywhere.
X phase gadgets are left unchanged under conjugation if the Z qubit is not a leg:
\[
  \includegraphics[height=0.08\textwidth]{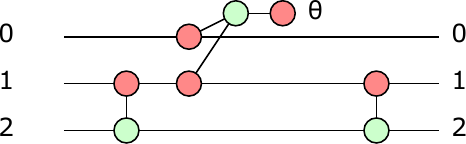}
  \hspace{1cm} \raisebox{0.03\textwidth}{=} \hspace{1cm}
  \includegraphics[height=0.08\textwidth]{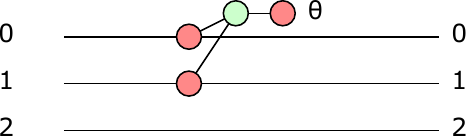}
\]
If the Z qubit is a leg and the X qubit is not a leg before conjugation, the X qubit becomes a leg after conjugation:
\[
  \includegraphics[height=0.08\textwidth]{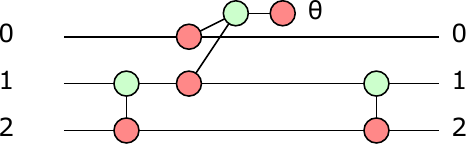}
  \hspace{1cm} \raisebox{0.03\textwidth}{=} \hspace{1cm}
  \includegraphics[height=0.08\textwidth]{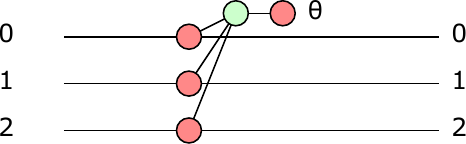}
\]
If the Z qubit is a leg and the X qubit is a leg before conjugation, the X qubit is no longer a leg after conjugation:
\[
  \includegraphics[height=0.08\textwidth]{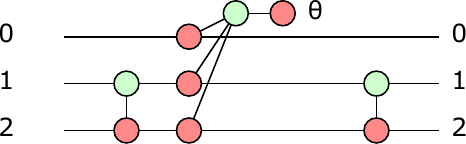}
  \hspace{1cm} \raisebox{0.03\textwidth}{=} \hspace{1cm}
  \includegraphics[height=0.08\textwidth]{svg-inkscape/gadg-cx-flip-example-1x_svg-raw.pdf}
\]
There are several different ways to obtain the conjugation rules above.
A simple one goes through the observation that conjugation commutes with matrix exponentiation:
\[
  A
  \left(
  \exp\left(i \theta \bigotimes_{q=0}^{n-1} G_q\right)
  \right)
  A^\dagger
  =
  \exp\left(i \theta\;
    A
    \left(
    \bigotimes_{q=0}^{n-1} G_q
    \right)
    A^\dagger
  \right)
\]
From this, it is enough to look at conjugation by CX for a handful of Pauli group elements:
\[
  \begin{array}{ccc}
    \text{cx}_{0,1}(\Z\otimes\I)\text{cx}_{0,1} = \Z\otimes\I
    &\;\;
    \text{cx}_{0,1}(\I\otimes\Z)\text{cx}_{0,1} = \Z\otimes\Z
    &\;\;
    \text{cx}_{0,1}(\Z\otimes\Z)\text{cx}_{0,1} = \I\otimes\Z\\
    \text{cx}_{0,1}(\I\otimes\X)\text{cx}_{0,1} = \I\otimes\X
    &\;\;
    \text{cx}_{0,1}(\X\otimes\I)\text{cx}_{0,1} = \X\otimes\X
    &\;\;
    \text{cx}_{0,1}(\X\otimes\X)\text{cx}_{0,1} = \X\otimes\I\\
  \end{array}
\]

\subsection{Mixed ZX Phase circuits}

A \emph{mixed ZX phase circuit} is, as the name suggests, a quantum circuit consisting of a mix of Z phase gadgets and X phase gadgets.
Examples of mixed ZX phase circuits include the ansatz\"{e} used by adiabatic quantum computation \cite{farhi2000quantum}, the quantum approximate optimisation algorithm \cite{farhi2014quantum}, as well as many quantum machine learning techniques \cite{yeung2020diagrammatic}.
The \texttt{pauliopt} library allows these circuits---called \emph{phase circuits}, for short---to be created directly from a sequence of phase gadgets with given angles and legs:
\begin{minted}{python}
circ = PhaseCircuit(3)
circ >>= Z(π/2)@{0,1}
circ >>= X(π)@{0}, Z(π/2)@{1}
circ >>= X(-π/4)@{1,2}
\end{minted}
\[
  \includegraphics[height=0.12\textwidth]{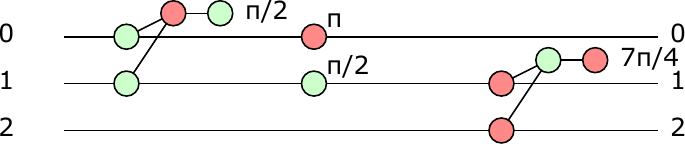}
\]
Utility methods are also made available to translate common gates into phase gadgets:
\begin{minted}{python}
circ = PhaseCircuit(3)
circ.ccz(0,1,2)
\end{minted}
\[
  \includegraphics[height=0.12\textwidth]{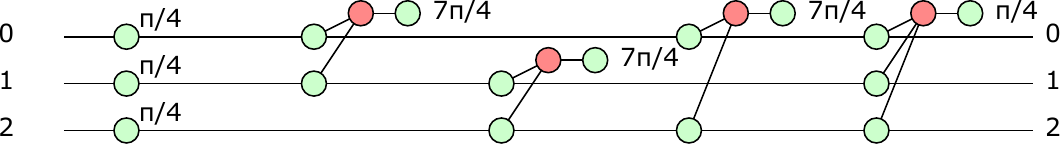}
\]
A generalisation of Euler's decomposition to $n$ qubits (technically, a consequence of Zassenhaus formula \cite{magnus1954,casas2012}) implies that every unitary on $n$ qubits can be expressed using a bounded number of Z and X phase gadgets, meaning that circuits of mixed ZX phase gadgets are universal.

\section{Annealing optimisation}

\subsection{Topology-aware cost of phase gadgets}

The optimisation method targets the CX count for phase gadget implementation on a given topology, which we refer to as its \emph{cost}.
The cost of a gadget relies on its implementation using a balanced tree of CX gates between its legs \cite{cowtan2019phase}, individually converted into double ladders of nearest-neighbour (NN) CX gates (without further simplification).
The balanced tree of minimum CX count is obtained using Prim's algorithm for minimum spanning trees, using the following weights for any two distinct leg qubits $q_i$ and $q_j$ in the phase gadget:
\[
  w(q_i, q_j) := 4d(q_i, q_j)-2
\]
where $d(q_i, q_j)$ is the graph distance between qubits $q_i$ and $q_j$ in the given topology.
For example, consider the following 4-legged gadget on a 3-by-3 grid topology:
\[
  \includegraphics[height=0.2\textwidth]{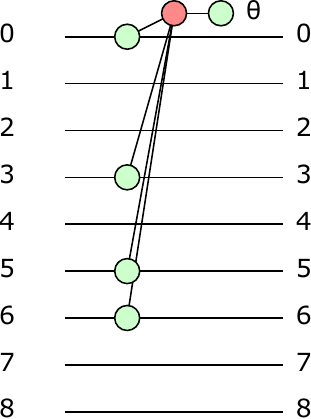}
  \hspace{2cm}
  \includegraphics[height=0.2\textwidth]{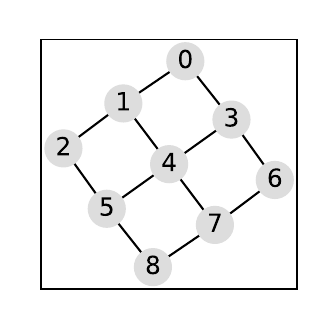}
\]
The minimum spanning tree implementation for this gadget uses two (adjoint) balanced trees of 3 CX gates each---between qubits $\{0, 3\}$ (NN CX count: 1), between qubits $\{3, 6\}$ (NN CX count: 1), and between qubits $\{3, 5\}$ (NN CX count: 3)---for a total implementation cost of 10 NN CX gates.
In this case, it is easy to see that every alternative would have been sub-optimal, because the only common nearest neighbour between qubits 0 and 6 on the grid is qubit 3: connecting both qubits to a qubit at distance 2 from both would have incurred a NN CX count of at least $3+3 = 6$, while connecting one qubit to a different nearest neighbour of the other (e.g. connecting 0 and 6 to 1) would have incurred a NN CX count of at least $5+1=6$, both already exceeding the optimal overall NN CX count of 5.
The following snippet produces the phase gadget in Qiskit \cite{Qiskit} using the above minimum spanning tree:
\begin{minted}{python}
from qiskit.circuit import QuantumCircuit, Parameter
cx_tree = QuantumCircuit(4)
cx_tree.cx(0, 2)
cx_tree.cx(1, 2)
cx_tree.cx(2, 3)
gadget = QuantumCircuit(8)
gadget.compose(cx_tree, qubits=(0,6,3,5), inplace=True)
gadget.rz(Parameter("θ"), 5)
gadget.compose(cx_tree.inverse(), qubits=(0,6,3,5), inplace=True)
\end{minted}
\[
\vspace{2mm}
  \raisebox{3.45mm}{
    \includegraphics[height=0.3\textwidth]{svg-inkscape/phase-gadg-cost-example-2_svg-raw.pdf}
  }
  \hspace{10mm}
  \raisebox{0.152\textwidth}{$\mapsto$}
  \hspace{8mm}
  \includegraphics[height=0.33\textwidth]{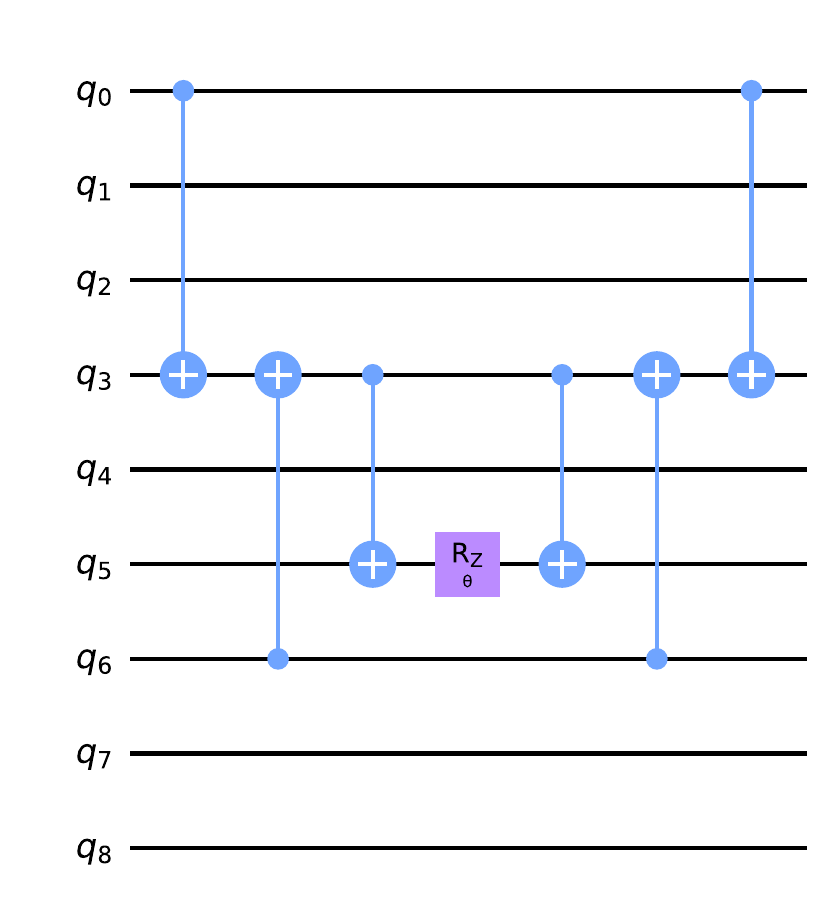}
\]

\subsection{Conjugation by CX circuits}

As discussed above, conjugation of mixed ZX phase circuits by CX circuits preserves the order and angles of the gadgets, but modifies their legs.
For adequate choices of nearest-neighbour CX circuits, the increase in CX count from the conjugating CX gates is counterbalanced by a decrease in implementation cost for the phase gadgets on a given topology, due to leg changes and reduction.
The overall 2-qubit gate count can then be used as a cost function for global optimisation methods (such as simulated annealing).
Here is example of a 10-gadget circuit on 4 qubits:
\[
  \includegraphics[width=\textwidth]{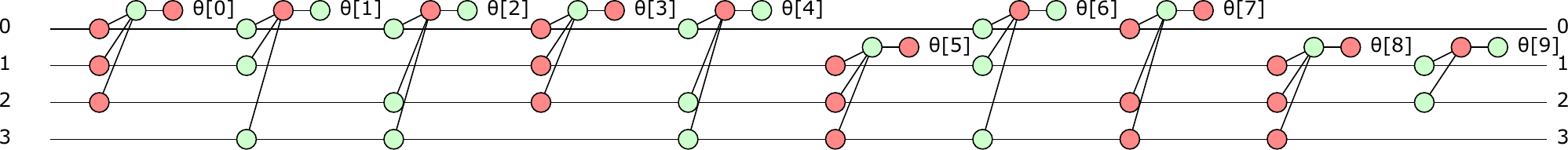}
\]
On a cycle topology, the circuit above has a CX count of 42.
Below is an equivalent circuit with a CX count of 14, obtained by conjugation from two layers of nearest-neighbour CX gates:
\[
  \includegraphics[width=0.8\textwidth]{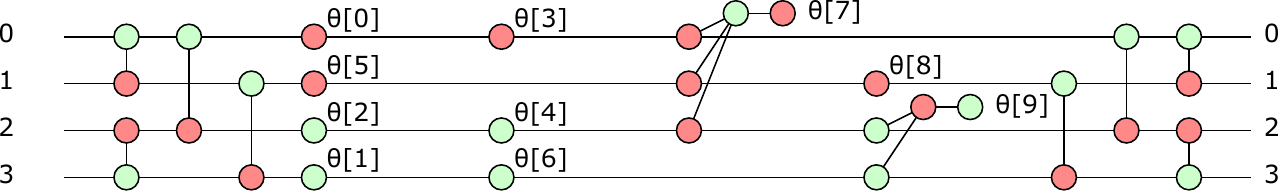}
\]
Here is a second example, of a 20-gadget circuit on 9 qubits:
\[
  \includegraphics[width=\textwidth]{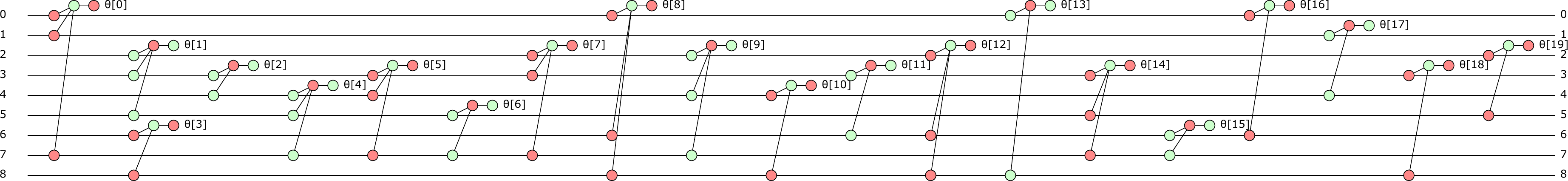}
\]
On a 3-by-3 grid topology, the circuit above has a CX count of 142.
Below is an equivalent circuit with a CX count of 126, obtained by conjugation from two layers of nearest-neighbour CX gates:
\[
  \includegraphics[width=\textwidth]{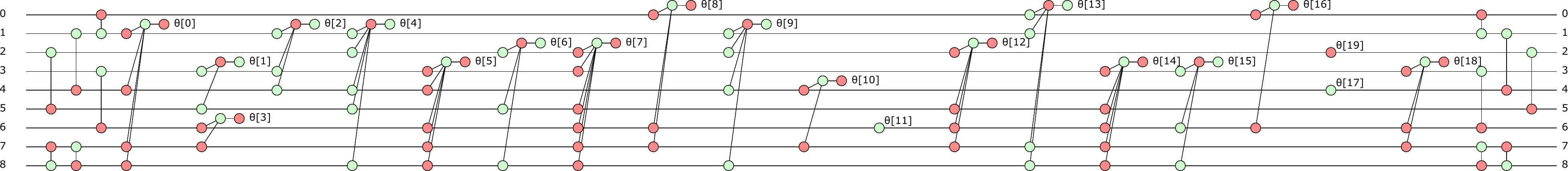}
\]
The two layers of conjugating CX gates for each one of the two example above, arranged on the corresponding qubit topologies, are displayed below:
\[
  \includegraphics[height=0.23\textwidth]{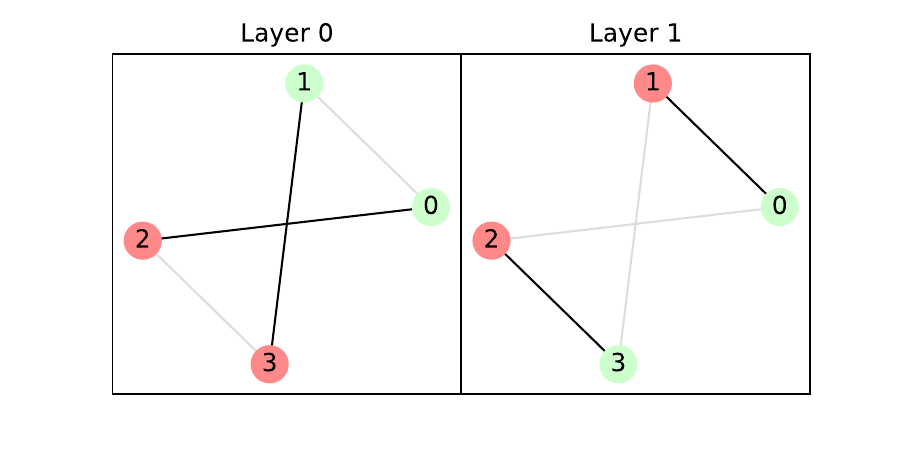}
  \includegraphics[height=0.23\textwidth]{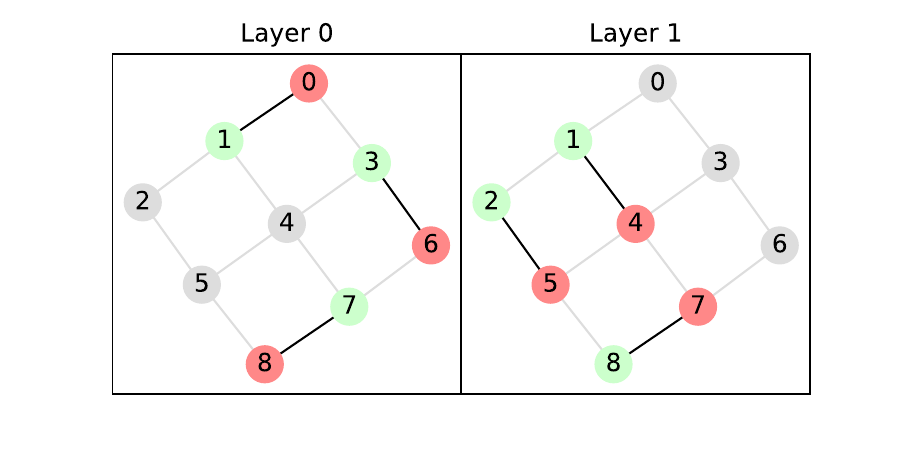}
\]

\subsection{Repeating layers}

A remarkable feature of our optimisation technique concerns its use with circuits composed of repeating layers.
For such circuits, it is enough to simplify an individual layer and then repeat it: the conjugating CX gates between repeating layers cancel out in pairs, leaving a single pair of conjugating CX block for the entire circuit.
\[
  \includegraphics[width=0.8\textwidth]{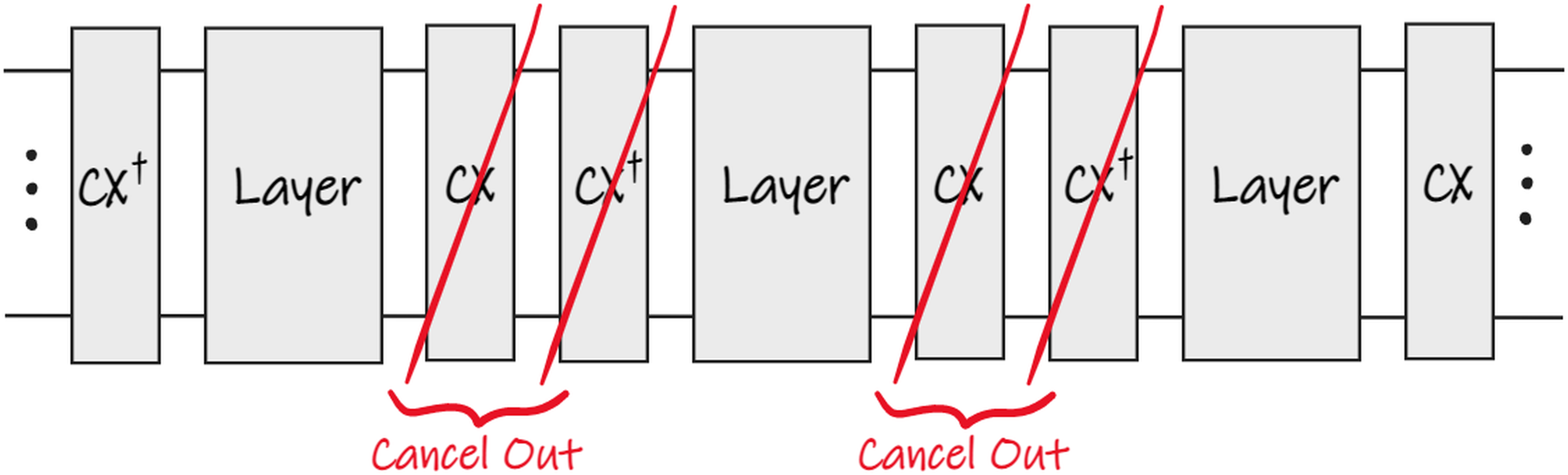}
\]
This means that the per-layer CX cost reduction is multiplied by the number of layers, while the CX cost for the conjugating blocks is fixed: as the number of layers grows, the cost of larger conjugating CX blocks is offset by the increased savings on the layers, at no additional computational expense for the optimisation itself.
As a concrete example, consider the following 4-gadget layer:
\[
  \includegraphics[height=0.14\textwidth]{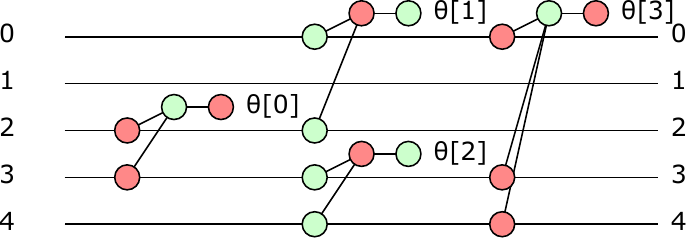}
\]
On a line topology with 5 qubits, the initial CX count for this layer is 22.
If optimisation is run for a single layer, the CX count is reduced by 18\%, from 22 to 18:
\[
  \includegraphics[height=0.14\textwidth]{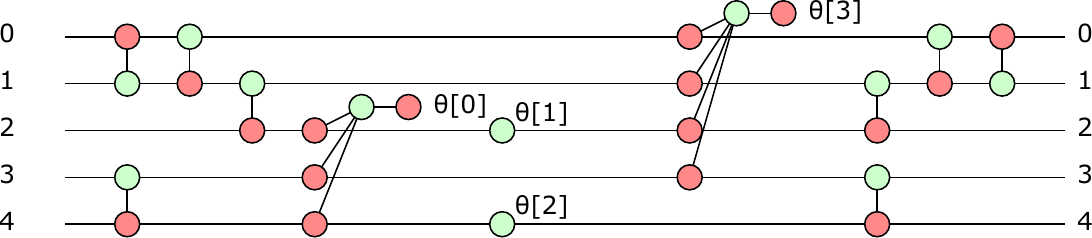}
\]
With 2 layers, the CX count is reduced by 36\%, from 44 to 28:
\[
  \includegraphics[height=0.14\textwidth]{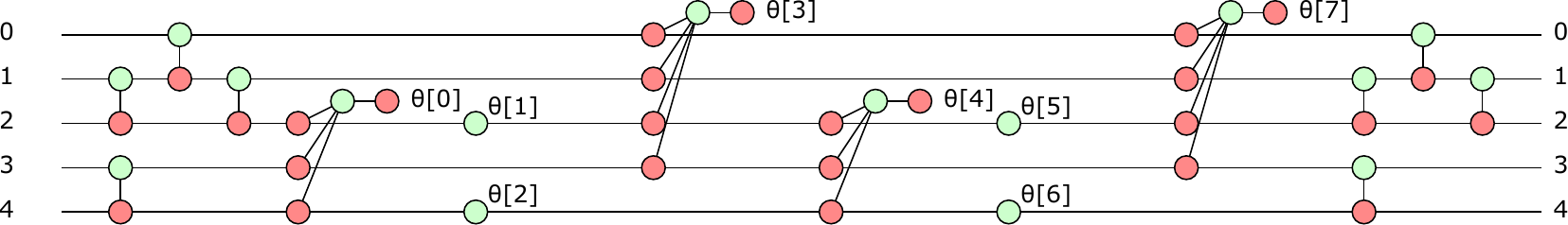}
\]
With 5 layers, the CX count is reduced by 47\%, from 110 to 58:
\[
  \includegraphics[width=\textwidth]{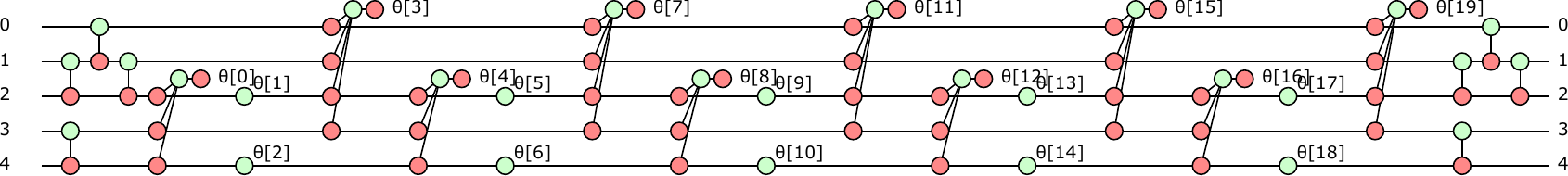}
\]
The 8 conjugating CX gates are independent of the number of layers, leaving 10 CX gates per simplified layer from an initial count of 22. In the limit of a large number of layers, the relative CX count reduction for this conjugating CX block approaches 12/22, or 54.5\%.

This feature of our optimisation technique makes it particularly interesting for applications in quantum machine learning (QML), adiabatic quantum computation and the quantum approximate optimisation algorithm (QAOA), where circuit ansatz\"{e} often feature the aforementioned repeating layers structure.
In particular, note that the technique is insensitive to the specific angles used in the repeated layers: the angles can be different, and even parametric (as indeed shown in the examples above).
This means that an ansatz circuit only need be optimised once, before assigning specific values to its parameter.
For a mixed ZX phase circuit formulation of several such ansatz\"{e}, we refer the reader to \cite{yeung2020diagrammatic}.

\subsection{Configuration space}

The configuration space explored by our optimisation algorithm on a given topology consists of all circuits of nearest-neighbour CX gates with a fixed number of layers, which we refer to as the \emph{CX blocks}.
At any given point in configuration space, i.e. at any such CX block $C$, the cost for a phase circuit $P$ is given by sum of:
\begin{itemize}
  \item the CX count for $C^\dagger$;
  \item the CX counts for the individual gadgets of $C^\dagger(P)$, computed on the given topology;
  \item the CX count for $C$;
\end{itemize}
where $C^\dagger(P) := C^\dagger \circ P \circ C$ is the phase circuit obtained from $P$ through conjugation by $C^\dagger$.
\[
  \includegraphics[width=0.7\textwidth]{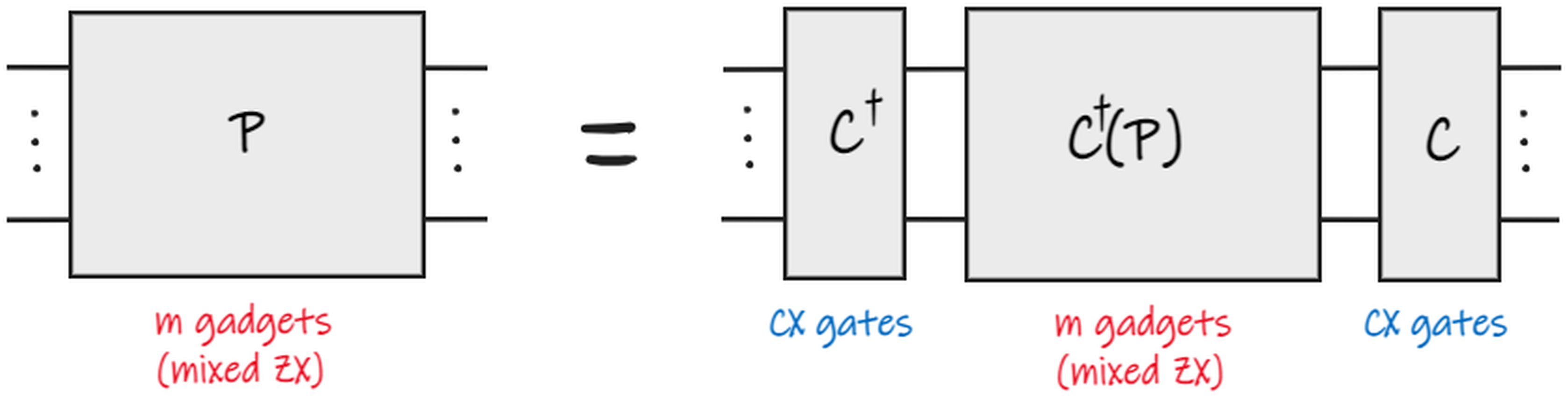}
\]
For fixed topology and number of layers, we say that a CX block $C'$ is obtained from another CX block $C$ by \emph{flipping} a CX gate $\text{cx}_{i, j}$ on a layer $l$ if either:
\begin{itemize}
  \item the gate $\text{cx}_{i, j}$ is present at layer $l$ in $C$, and $C'$ is obtained from $C$ exactly by removing $\text{cx}_{i, j}$ from layer $l$;
  \item there is no gate at layer $l$ in $C$ which is incident to either qubit $i$ or qubit $j$ (or both), and $C'$ is obtained from $C$ exactly by adding  $\text{cx}_{i, j}$ to layer $l$.
\end{itemize}
We write $C \stackrel{l; i, j}{\longrightarrow} C'$ to denote this fact; because CX gates are self-inverse, this is equivalent to $C \stackrel{l; i, j}{\longleftarrow} C'$.
We consider two CX blocks to be nearest neighbours in configuration space if one can be obtained from the other by a single CX gate flip: this endows the configuration space with the structure of an undirected graph, where edges are labelled by the triples $(l; i, j)$ describing the CX gate flips.
Our optimisation algorithm performs a random walk on this graph, starting from the empty CX block and attempting to (globally) minimise the cost for a given phase circuit.

\subsection{Simulated Annealing}

Simulated annealing is a global optimisation method which explores a discrete problem-dependent configuration space $\mathcal{X}$, in an attempt to find an approximate optimum for a an arbitrary cost function $f$.
At each step of the optimisation, the algorithm sits at some configuration $x \in \mathcal{X}$ and selects a uniformly random ``direction'' to explore; that is, it samples a uniformly random $x' \in N(x)$, where $N(x) \subseteq \mathcal{X}\backslash\{x\}$ a bounded set of nearest neighbours of $x$ in configuration space.
The algorithm the computes $f(x')$ and compares it to $f(x)$: if $f(x') \leq f(x)$, then the algorithm moves to $x'$ deterministically; otherwise, it moves to $x'$ with a probability which decreases exponentially in the difference $\Delta := f(x')-f(x)$:
\[
\text{Prob}\left(x \rightarrow x'\right)
=
\exp\left(-\frac{\Delta\log(2)}{t}\right)
\]
The parameter $t > 0$ is known as the \emph{temperature}: it decreases from an initial value $t_{0}$ to a final value $t_N \approx 0$ over $N$ iterations, in a way specified by the ``temperature schedule''.
The temperature schedule for the annealing determines the probability of accepting changes which increase the cost function: an increase of $t\log_2(1/p)$ has probability $p$ of being accepted.
At the start of the annealing, when the temperature is high, the random walk is allowed to explore the configuration space, even at the cost of an increase in cost.
As iterations progress and the temperature is lowered, the random walk progressively favours those steps which decrease the cost.
Towards the end, the probability of accepting steps which increase the cost becomes vanishingly small, and the algorithm is reduced to a greedy optimisation.

\subsection{Simulated Annealing for Mixed ZX Phase Circuit Optimisation}

For our problem, the simulated annealing configuration space consists of a fixed number of layers of NN CX gates on the given topology, with nearest neighbours in configuration space determined by CX gate flips in any one of the layers.
Given a mixed ZX phase circuit, a topology and a fixed number of layers for the CX blocks, we construct an optimized circuit container; initially, this consists of the phase circuit conjugated by empty CX blocks.
To minimise the cost of over the space of CX blocks, we perform a random walk using simulated annealing \cite{kirkpatrick671}, for a given number of steps/iterations and following a given temperature schedule.
At the end of the annealing, the optimized circuit contains a simplified phase circuit conjugated by (typically) non-empty CX blocks.
Every iteration of our simulated annealing algorithm proceeds as follows:
\begin{enumerate}
  \item We obtain the current temperature $t$ from the temperature schedule using the current iteration number (e.g. linearly interpolating from initial temperature to final temperature).
  \item We select a random CX gate flip for the current block, i.e. select a random neighbour.
  \item We flip the selected CX gate (see Subsection \ref{section:flipping-gates} below).
  \item We calculate the CX count $c_{new}$ for the new circuit and compare it to the previous CX count $c_{old}$, by computing $\Delta := c_{new} - c_{old}$ and sampling a uniformly random number $r \in [0, 1]$:
    \begin{itemize}
      \item if $r \leq \exp\left(-\frac{\Delta\log(2)}{t}\right)$, and in particular if $c_{new} \leq c_{old}$, the move is accepted and we go to the next iteration from the new CX block;
      \item otherwise, the move is undone---by flipping the selected CX gate again---and we go the next iteration from the old CX block.
    \end{itemize}
\end{enumerate}
Our selection of temperature schedule depends on the shape of the temperature curve (linear, geometric, reciprocal or logarithmic) and two parameters: the amount of time we want to spend exploring the configuration space at the start of the annealing and the amount of time we want to spend performing greedy optimisation at the end of the annealing.
As a convention, we take the temperature value $t=2m+2$ (where $m$ is the number of phase gadgets) to denote the end of the exploration phase and the temperature value $t=2$ to denote the start of the greedy optimisation phase.
Because the CX gates are nearest-neighbour, $2m+2$ is the maximum possible increase in CX count from a single CX gate flip: 2 CX gates added plus a leg added to all gadgets, at a worst-case CX count increase of 2 per gadget. On the other hand, $2$ is the minimum possible (positive) increase: CX count always changes by multiples of 2.
This defines the exploration phase as the time where every CX gate flip has a probability higher than $50\%$ of being accepted, and the greedy optimisation phase as the time where every increase in CX count has a probability lower than $50\%$ of being accepted.
As an example, below is a sketch of a linear temperature schedule over $N$ iterations:
\[
  \includegraphics[width=0.5\textwidth]{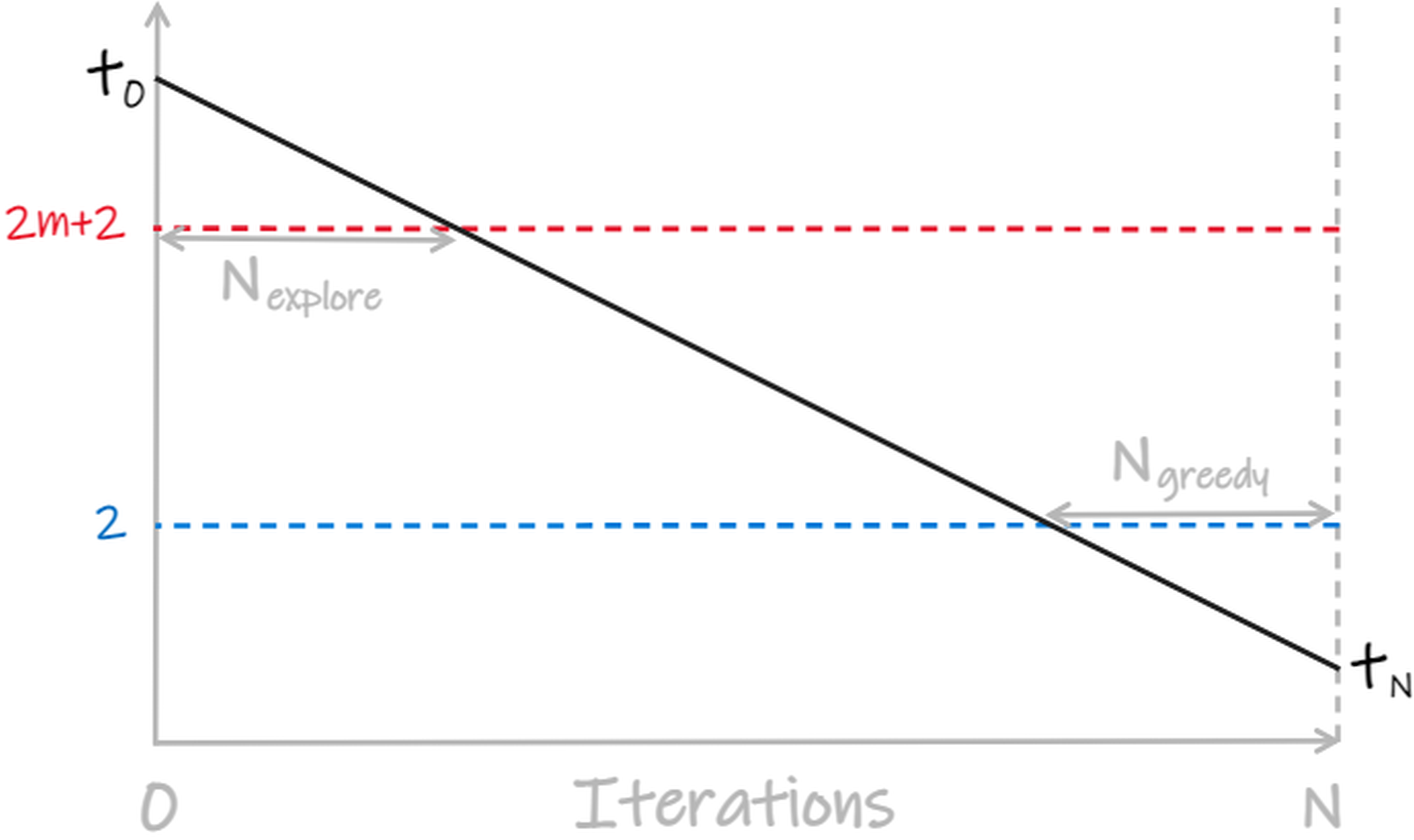}
\]

\subsection{Flipping CX gates}
\label{section:flipping-gates}

Because simulated annealing requires thousands of iterations, optimised execution of CX gate flips is paramount.
Naively, one could simply maintain the current CX block $C$ and compute the conjugated circuit from scratch at every iteration: this requires conjugating every gadget in the phase circuit by every CX gate in the CX block, even though the CX block itself has only changed by one gate flip.
In this work, we adopt a different strategy, where we maintain both the current CX block $C$ and the current conjugated phase circuit $C^\dagger(P)$.
Imagine we wish to perform a gate flip $C \stackrel{l; i, j}{\longrightarrow} C'$, e.g. performing the flip $C \stackrel{3; 5, 4}{\longrightarrow} C'$ boxed in dark blue below.
\[
  \includegraphics[width=0.6\textwidth]{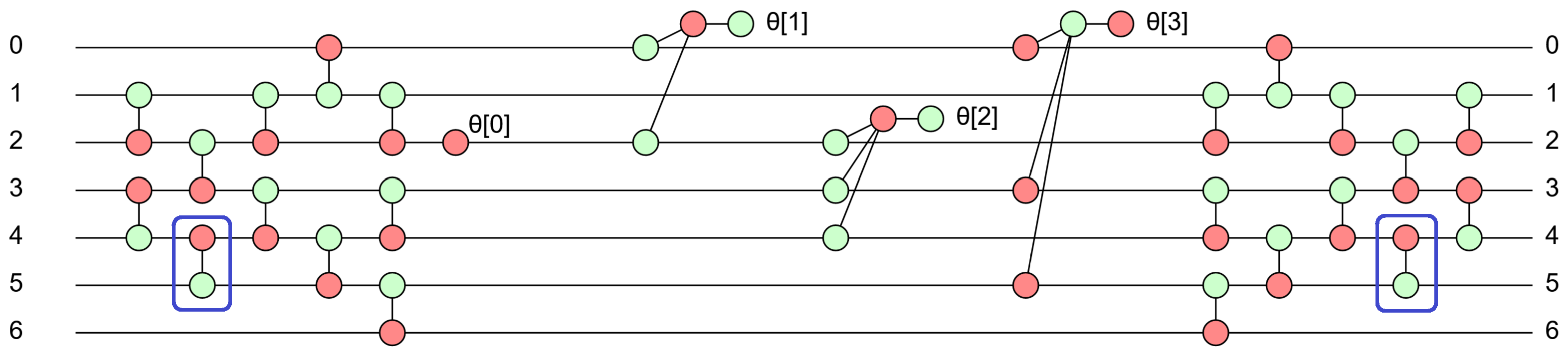}
\]
To perform the flip, we go through the following 4 steps:
\begin{enumerate}
  \item Starting from layer $l-1$ and moving inwards towards layer $0$, we create the (partially ordered) set $(l; i, j)\!\!\downarrow$ of all CX gates that are strictly in the ``past'' of gate $\text{cx}_{i, j}$ at layer $l$. These are boxed and highlighted below.
  \[
    \includegraphics[width=0.6\textwidth]{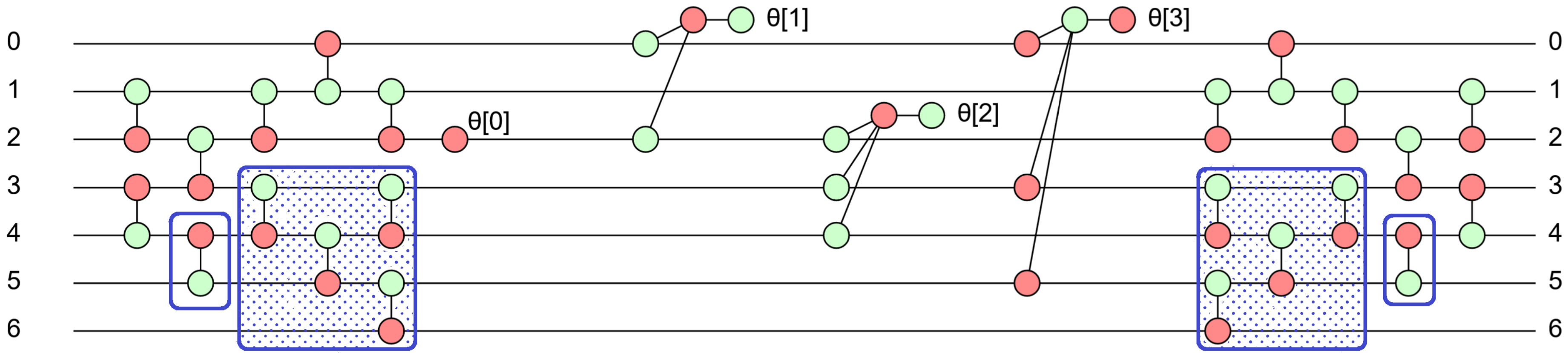}
  \]
  \item We undo all the CX gates in $(l; i, j)\!\!\downarrow$, conjugating the phase circuit by each one, working our way outwards from layer $0$ (closest to the phase circuit) up to layer $l-1$ (just below the gate we'll flip).
  \[
    \includegraphics[width=0.6\textwidth]{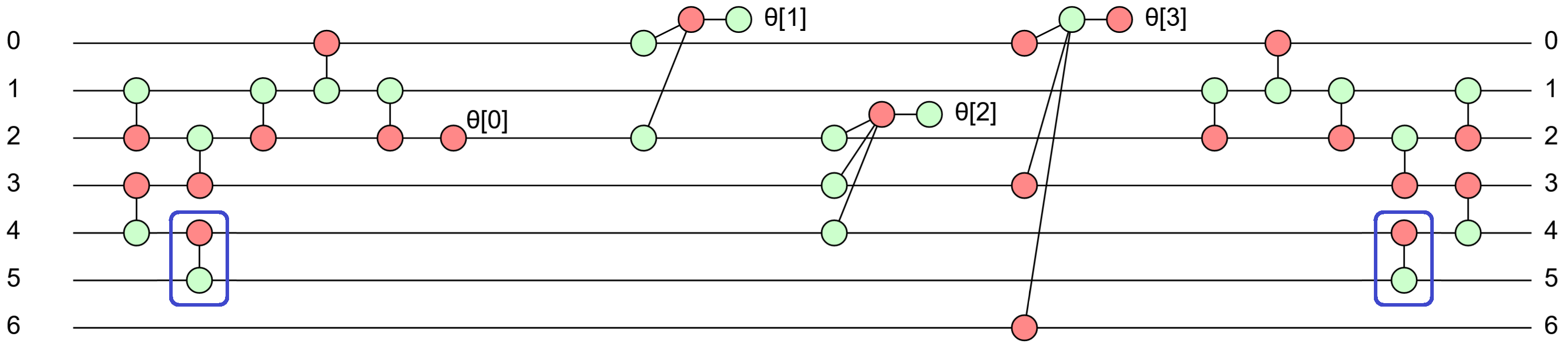}
  \]
  \item We flip gate $\text{cx}_{i, j}$ at layer $l$, conjugating the phase circuit by $\text{cx}_{i, j}$ (adding $\text{cx}_{i, j}$ to layer $l$ if it wasn't there, removing it from layer $l$ if it was). Below, we remove $\text{cx}_{5, 4}$ from layer $3$.
  \[
    \includegraphics[width=0.6\textwidth]{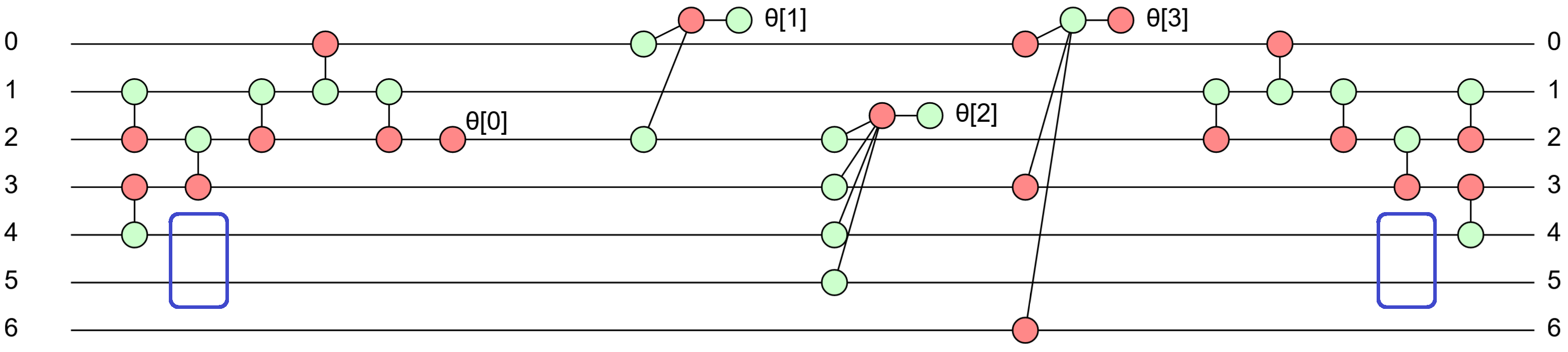}
  \]
  \item We redo all the CX gates in $(l; i, j)\!\!\downarrow$, conjugating the phase circuit by each one, this time working our way inwards from layer $l-1$ down to layer $0$.
  \[
    \includegraphics[width=0.6\textwidth]{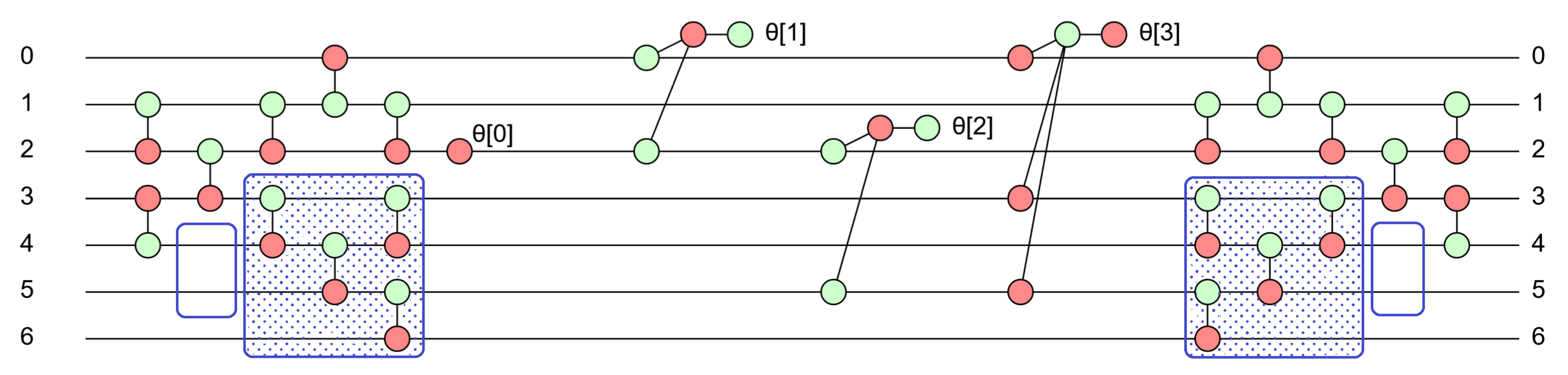}
  \]
\end{enumerate}
To define the poset $(l; i, j)\!\!\downarrow$ of past gates formally, consider the partial order on the gates $(l; x, y)$ of the CX block obtained by transitive-reflexive closure of the following relation:
\[
(l_1; x_1, y_1) < (l_2; x_2, y_2)
\Leftrightarrow
l_1 < l_2
\text{ and }
\{x_1, y_1\} \cap \{x_2, y_2\} \neq \emptyset
\]
The poset $(l; i, j)\!\!\downarrow$ consists of all gates strictly less than $(l; i, j)$ in this partial order.

\section{Performance Evaluation}

\subsection{Random mixed ZX phase circuits}

To evaluate the performance of our technique, we test it on random mixed ZX phase circuits with a varying number 2- and 3-legged gadgets, on 16, 25 and 36 qubits arranged in a square grid.
\vspace{-4mm}
\[
  \includegraphics[height=0.18\textwidth]{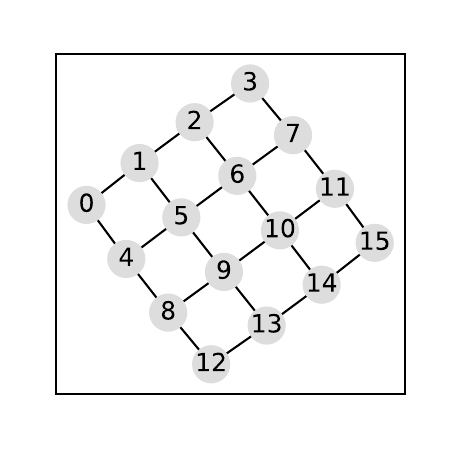}
  \hspace{20mm}
  \includegraphics[height=0.18\textwidth]{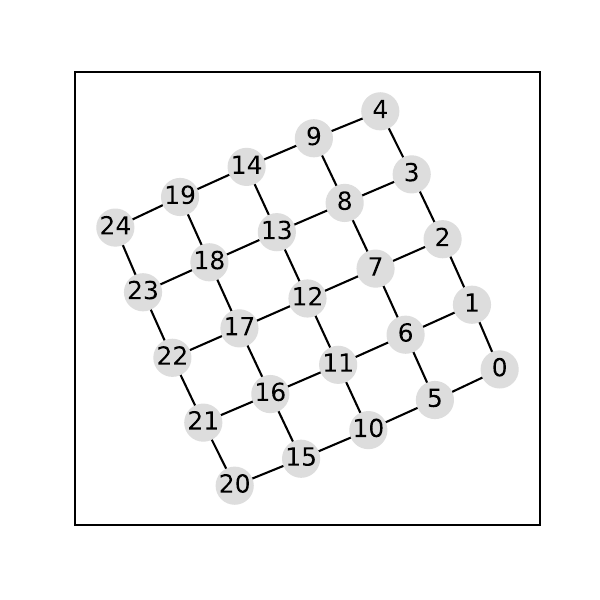}
  \hspace{20mm}
  \includegraphics[height=0.18\textwidth]{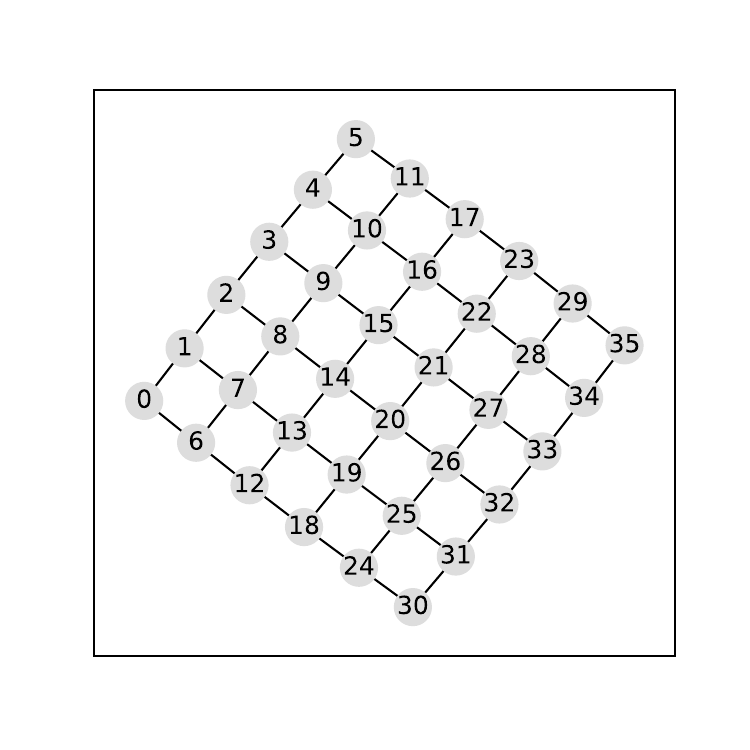}
  \vspace{-5mm}
\]
Our random model samples a fixed number of gadgets \texttt{num\_gadgets} on a fixed number of qubits \texttt{num\_qubits}, selecting legs in a independent, identically distributed way for each gadget.
Legs for a gadget are sampled by first selecting the number of legs uniformly randomly in a given range (from \texttt{min\_legs} to \texttt{max\_legs}, both included), and then selecting a uniformly random subset of qubits with the desired cardinality.
\begin{minted}{python}
PhaseCircuit.random(num_qubits, num_gadgets, *, min_legs, max_legs)
\end{minted}
Below is an example of a random circuit of 10 gadgets on 6 qubits, with between 1 and 3 legs:
\[
  \includegraphics[width=0.8\textwidth]{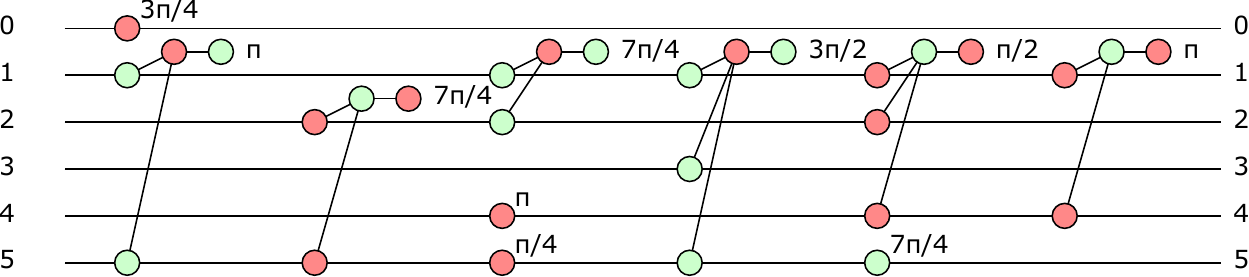}
\]
We can also make our random circuits purely parametric, but this doesn't impact our benchmarking.

\subsection{Internal representation of mixed ZX phase circuits}

The internal representation of phase circuits in our technique is optimised for execution of the the annealing algorithm.
A circuit ($n$ qubits, $m_z$ Z gadgets and $m_x$ X gadgets, in any order) is broken down into the following components:
\begin{itemize}
  \item a pair of binary matrices (an $n \times m_z$ matrix $L_z$ for Z gadgets and an $n \times m_x$ matrix $L_x$ for X gadgets) encoding the positions of the legs;
  \item a pair of lists (length $m_z$ for Z gadgets and length $m_x$ for X gadgets) mapping the columns of each matrix to the position of the corresponding gadget in the circuit;
  \item a list (length $m_z + m_x$) of angles for the gadgets (concrete or parametric).
\end{itemize}
The annealing algorithm operates on gadget legs only---i.e. on the the two matrices---without any alteration to the original gadget order and angles.
The algorithm further acts on the columns of each matrix independently, leaving ample scope for caching (implemented) and GPU parallelisation (not yet implemented).
As an example, consider the following circuit of $m_z+m_x=5$ gadgets on $n=3$ qubits:
\[
  \includegraphics[height=0.1\textwidth]{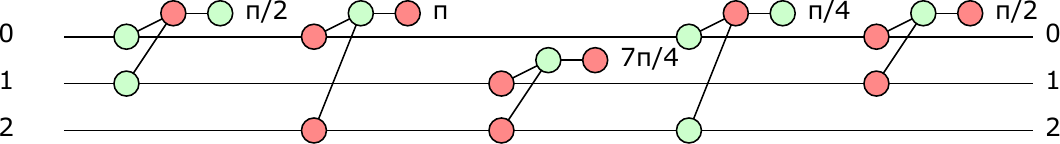}
\]
The matrices for this circuit are as follows:
\[
L_z =
\left(
\begin{array}{cc}
  1 & 1 \\
  1 & 0 \\
  0 & 1
\end{array}
\right)
\hspace{3cm}
L_x =
\left(
\begin{array}{ccc}
  1 & 0 & 1\\
  0 & 1 & 1\\
  1 & 1 & 0
\end{array}
\right)
\]
The lists mapping columns to gadgets are $(0, 3)$ for $L_z$ and $(1, 2, 4)$ for $L_x$. The list of angles for the gadgets is $(\pi/2, \pi, 7\pi/4, \pi/4, \pi/2)$.

\subsection{Benchmarking Results}

In all benchmarks below, we proceed as follows:
\begin{enumerate}
  \item We generate 50 random mixed ZX phase circuits, for various combinations of hyper-parameters---number of qubits, number of gadgets per layer, initial temperature, schedule type---randomly sampling between 2-legged and 3-legged gadgets. The number of CX block layers is fixed to 3.
  \item We run annealing optimisation with varying number of iterations, from 100 to 5000, in increments of 100. The number of circuit layer repetitions is fixed to 5.
  \item We look at the reduction in NN CX count---in average or in distribution---from each original random circuit to the corresponding optimised circuit, relative to the NN CX count of the original circuit. The NN CX count is computed using Prim's algorithm for minimum spanning trees.
\end{enumerate}
To start with, we look at the performance for four different initial temperatures $t_0=1, 5, 10, 20$ and two annealing schedules (linear and geometric).
Below is the average CX count reduction over all runs for the linear annealing schedule at the four initial temperatures.
\[
  \includegraphics[width=\textwidth]{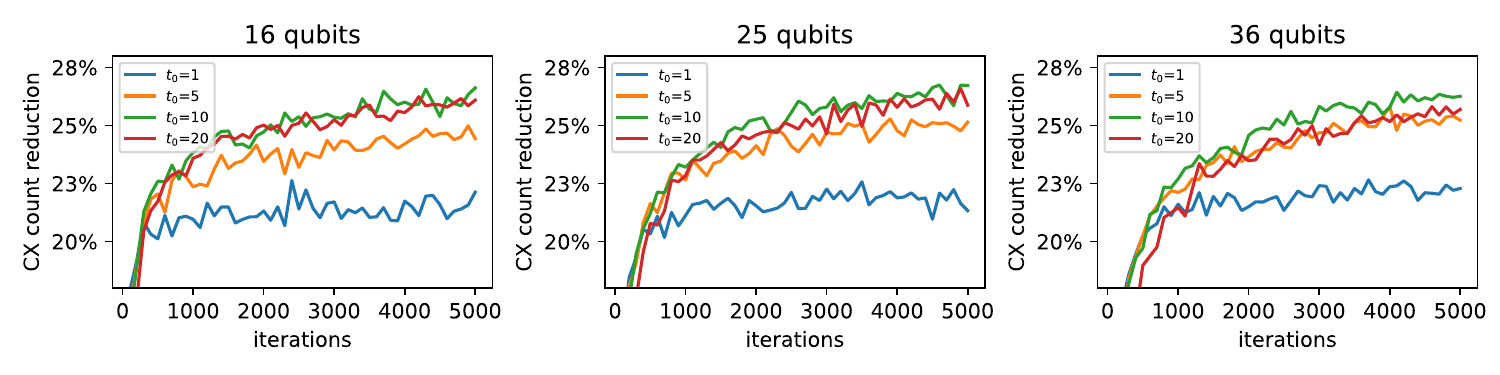}
\]
Below is the average CX count reduction over all runs for the geometric annealing schedule at the four initial temperatures. Because of the inferior average performance of other combinations, we restrict our attention to linearly annealed runs with $t_0=10$ for the remainder of this section.
\[
  \includegraphics[width=\textwidth]{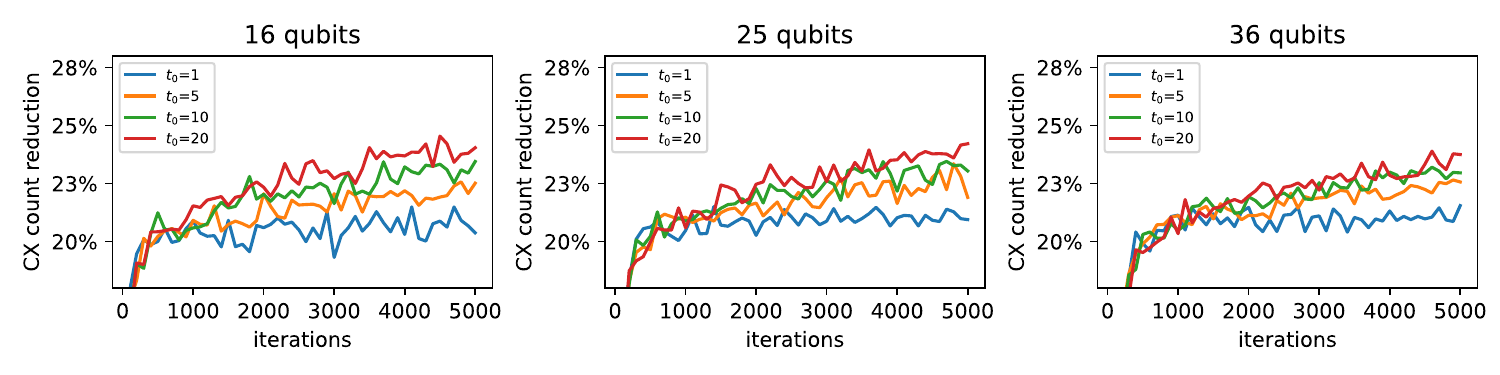}
\]
For our optimisation, we always use 5 layer repetitions for the phase circuit and 3 layers for the conjugating CX blocks.
Below is the average CX count reduction---as a percentage of the initial CX count---for varying number $m$ of phase gadgets per circuit layer, ranging from 10 to 35 in increments of 5.
\[
  \includegraphics[width=\textwidth]{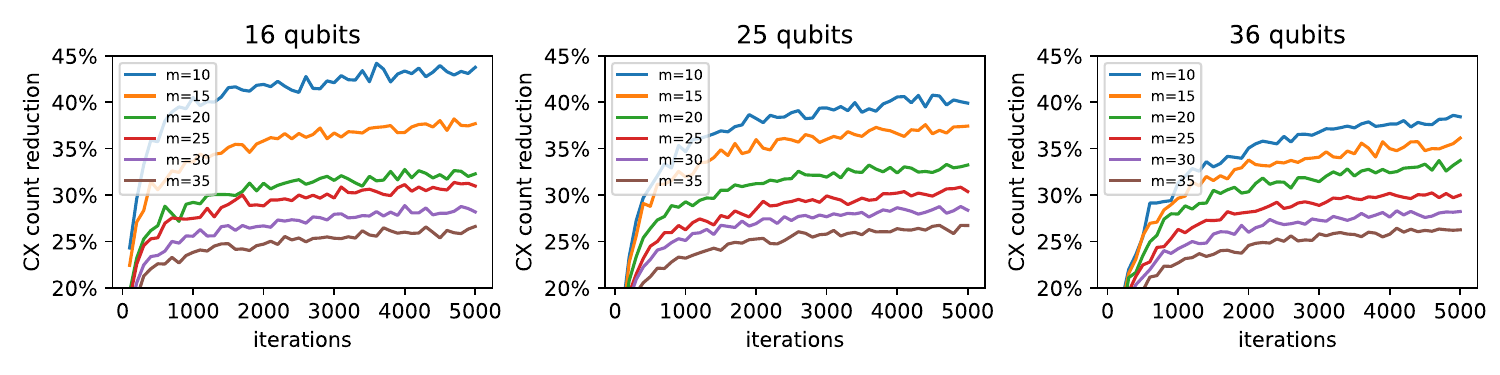}
\]
The average CX count reduction degrades as the number of gadgets/layer increases, decreasing by approximately 0.5\% per additional gadget on 36 qubits.
\[
  \includegraphics[width=\textwidth]{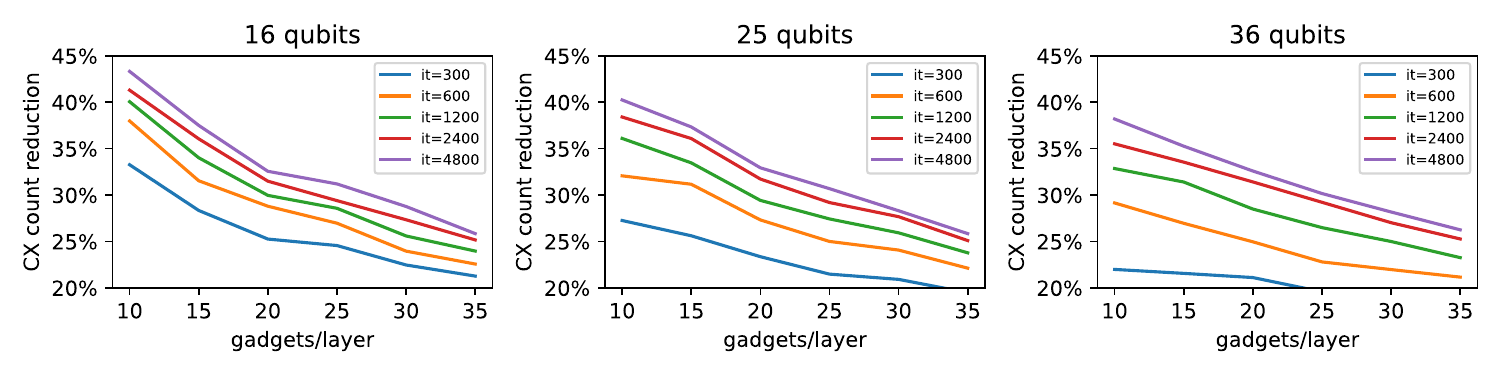}
\]
This behaviour is roughly linear when the number of gadgets/layer is small compared to the number of qubits, but approaches a horizontal asymptote as the number of gadgets/layer increases.
The following graphs showcase this behaviour over smaller grid topologies ($3\times3$, $3\times4$, $4\times4$ and $4\times5$ respectively).
\[
  \includegraphics[width=0.92\textwidth]{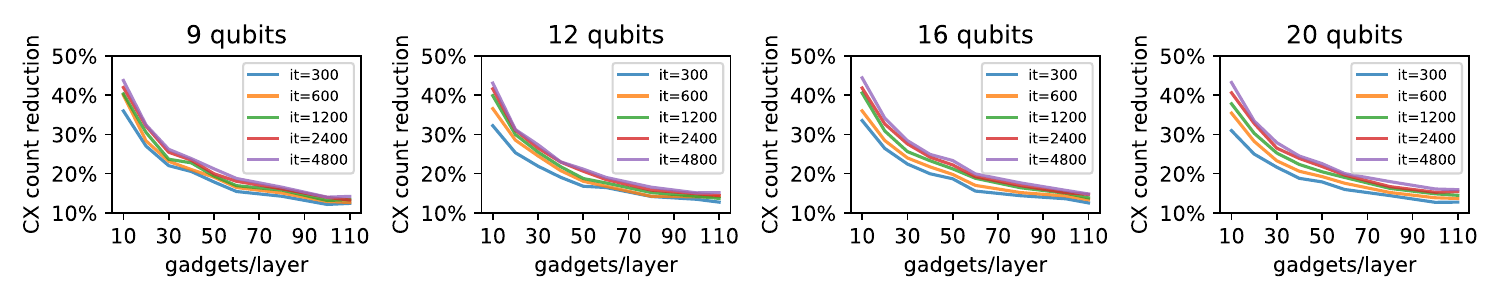}
\]
The optimised flipping of CX gates allows us to avoid recomputing the whole conjugated circuit every time.
We also implement some basic caching techniques for gadget legs, and use a modified version of Prim's algorithm to efficiently compute the topologically-aware CX count for each gadget.
Our Python implementation is reasonably efficient, taking between $400\mu s$ and $800\mu s$ per iteration.
\[
  \includegraphics[width=\textwidth]{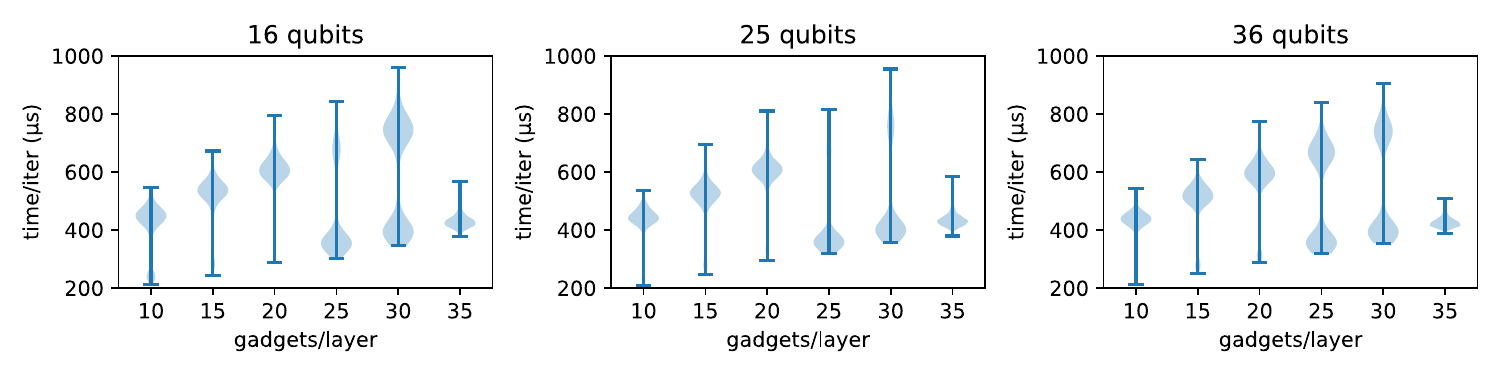}
\]
Over the full range of iterations we considered, from 100 to 5000, our experiments never took more than $4s$, with the vast majority under $3s$.
\[
  \includegraphics[width=\textwidth]{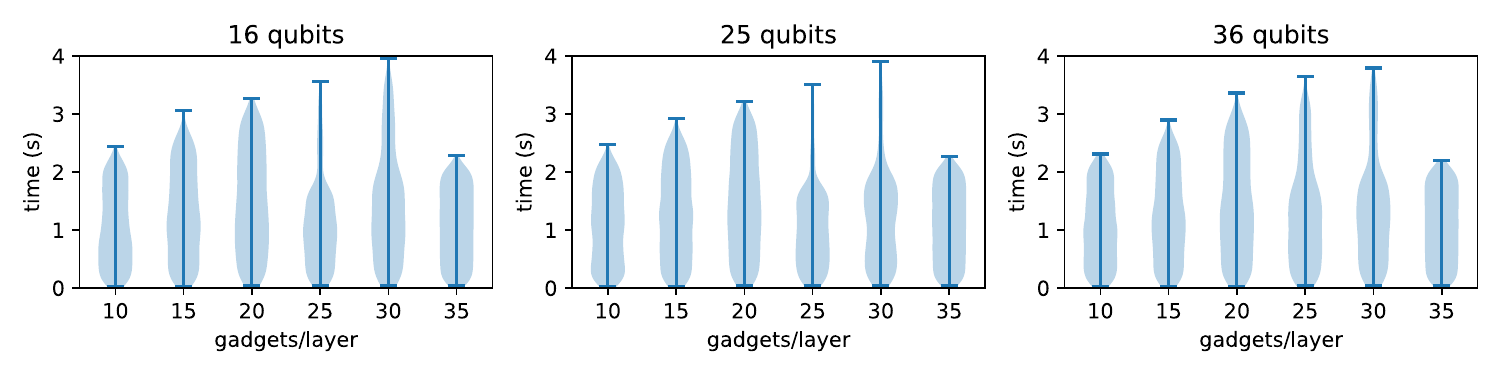}
\]
To evaluate the performance of our approach under strict time constraints, we restrict our attention to the runs that finished under $1s$, using almost all the time they were allowed (in the 90th to 100th percentile within runs under $1s$ for the same number of qubits and gadget/layers).
We observe that the average CX count reduction decreases as the number of gadgets/layer increases, and that the distribution narrows.
\[
  \includegraphics[width=\textwidth]{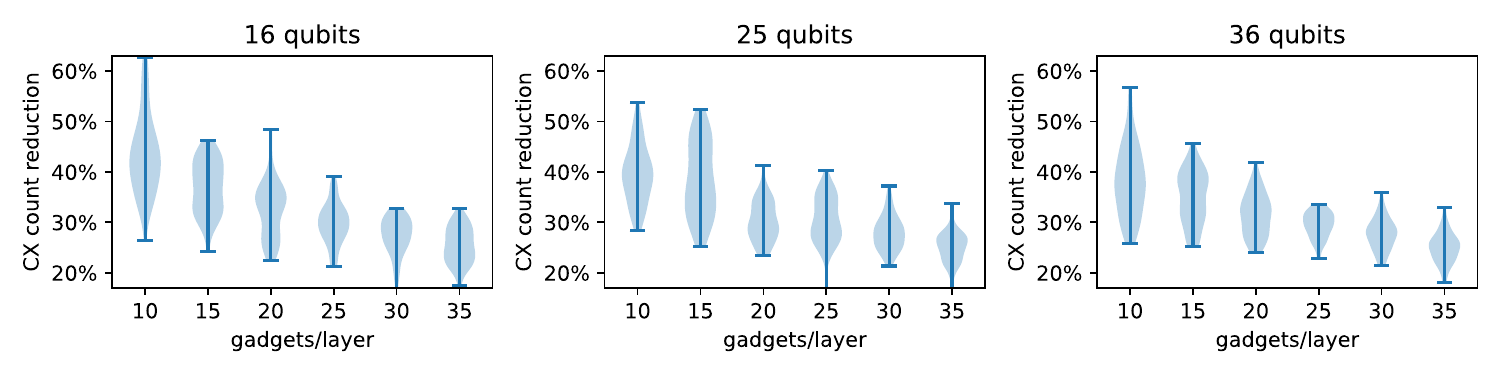}
\]
To further evaluate the effect of time on our optimisation, we restrict our attention to runs at 30 gadgets/layer, partitioned into time bins with $0.5s$ resolution.
The average CX count reduction improves only slightly as more time is allowed, indicating diminishing returns for progressively higher number of iterations.
However, we observe a marked improvement of worst-case CX count reduction from anneals under $0.5s$ to anneals up to $1s$, with only marginal improvement after that point.
\[
  \includegraphics[width=\textwidth]{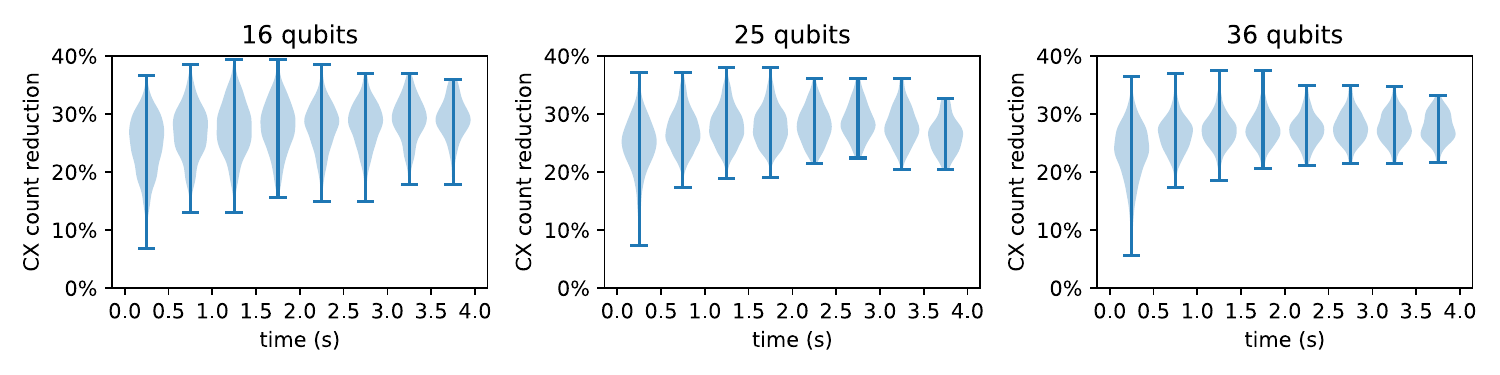}
\]

\section{Future Work}

Our first avenue for future work concerns the annealing itself: rather than using simulated annealing on classical hardware, we have started working on a formulation of the problem suitable for execution on D-Wave quantum annealers.
This would allow our exploration of the space of configurations to efficiently scale to much larger examples, involving hundreds of gadgets and qubits.
It would also make our technique a ``quantum optimisation of quantum circuits'', which is fun to say.
Thanks to QPL 2022 Anonymous Reviewer \#1 for pointing us to \cite{andrew2014ising}, which provides an Ising formulation for the minimum spanning tree objective.

Our second avenue for future work concerns the scope and method of optimisation: instead of limiting ourselves to mixed phase gadgets and random CX circuits, we have have generalised our observation to arbitrary layers of Pauli gadgets using random Clifford circuits.
We are in the process of implementing this technique and will report on its performance in future work.
We also plan to incorporate commutation between phase gadgets in different bases to further improve performance. 

We will benchmark our techniques explicitly against previous literature, using realistic circuit families from quantum chemistry, adiabatic quantum computation, quantum approximate optimisation and quantum machine learning.

\section*{Acknowledgements}

The authors would like to acknowledge financial support by Hashberg Ltd for the execution of numerical experiments.
The authors would like to thank the three anonymous QPL 2022 reviewers for their very helpful comments and suggestions.

\bibliographystyle{eptcs}
\bibliography{bibliography}

\end{document}